\documentclass{article}

\usepackage[utf8]{inputenc}
\usepackage[T1]{fontenc}   
\usepackage{hyperref}      
\usepackage{url}           
\usepackage{booktabs}      
\usepackage{amsfonts}      
\usepackage{amsmath}       
\usepackage{bm} 
\usepackage{bbm}
\usepackage{nicefrac}   
\usepackage{microtype}  
\usepackage{graphicx}
\usepackage{cleveref}
\usepackage{subcaption}
\usepackage[font=footnotesize]{caption}

\usepackage[preprint]{spconf}
\toappear{
\begin{tabular}{l}
\hspace*{-5.5mm}\footnotesize \copyright\ IEEE 2024. Personal use of this material is permitted.  Permission from IEEE must be obtained for all other uses, in any current or future media, including \\[-1.4mm]
\hspace*{-5.5mm}\footnotesize \phantom{\copyright } reprinting/republishing this material for advertising or promotional purposes, creating new collective works, for resale or redistribution to servers or lists, or \\[-1mm]
\hspace*{-5.5mm}\footnotesize \phantom{\copyright } reuse of any copyrighted component of this work in other works.
\end{tabular}
}

\usepackage{stackengine}

\usepackage{spconf,amsmath,graphicx}

\usepackage{enumitem}
\setlist{nosep, leftmargin=14pt}

\let\P\relax
\DeclareMathOperator{\P}{\mathbb{P}}

\DeclareMathOperator{\expreg}{\mathnormal{g}}
\DeclareMathOperator{\params}{\bm{\theta}}
\DeclareMathOperator{\mparams}{\bm{\phi}}
\DeclareMathOperator{\n0}{\mathnormal{n}_\mathrm{0}}

\DeclareMathOperator*{\argmax}{\arg\,\max}
\DeclareMathOperator{\U}{\mathcal{U}}
\DeclareMathOperator{\N}{\mathbb{N}}
\DeclareMathOperator{\R}{\mathbb{R}}
\DeclareMathOperator{\obs}{\mathbf{Y}}
\DeclareMathOperator{\cct}{\mu_{\mathnormal{C}}}
\DeclareMathOperator{\tc}{{\mathnormal{t_C}}}
\DeclareMathOperator{\likeli}{\mathcal{L}}
\DeclareMathOperator{\normal}{\mathcal{N}}

\DeclareMathOperator{\design}{\mathbf{X}}
\DeclareMathOperator{\Nexp}{\mathnormal{N}_{\bm{\theta}}}

\DeclareMathOperator{\Nmulti}{\mathnormal{N}_{\bm{\phi}}}
\DeclareMathOperator{\mNmulti}{\mathnormal{m}_{\bm{\phi}}}
\DeclareMathOperator{\vNmulti}{\mathnormal{v}_{\bm{\phi}}}
\newcommand{\tNmulti}[1]{\mathnormal{N}_{\bm{\phi}}^{#1}}

\newcommand{\Lp}{$\mathnormal{L}^{\mathrm{1}}$~}

\newcommand{\multistage}{\texttt{multi-stage}}
\newcommand{\GTsub}{\texttt{GT} \texttt{sub}}
\newcommand{\AT}{\texttt{ActiveTrack}}
\newcommand{\KITGE}{\texttt{KIT-GE}}
\newcommand{\MUCZ}{\texttt{MU-CZ}}
\newcommand{\tracking}{\AT, \KITGE{} and \MUCZ}

\DeclareMathOperator{\E}{\mathbb{E}}
\DeclareMathOperator{\Var}{\mathrm{Var}}

\newcommand\numberthis{\addtocounter{equation}{1}\tag{\theequation}}

\makeatletter
\renewcommand\@makefntext[1]{%
\setlength\parindent{1em}%
\noindent
\mbox{$^{\@thefnmark}$~}{#1}}
\makeatother

\title{Robust Approximate Characterization of Single-Cell Heterogeneity in Microbial Growth}
 \name{
    Richard D. Paul$^{\star}$ \qquad 
    Johannes Seiffarth$^{\dagger,\ddagger}$ \qquad 
    Hanno Scharr$^{\star}$ \qquad 
    Katharina N\"oh$^{\dagger}$}

 \address{
    $^{\star}$ IAS-8: Data Analytics and Machine Learning, Forschungszentrum Jülich GmbH, Jülich, Germany. \\
    $^{\dagger}$ IBG-1: Biotechnology, Forschungszentrum Jülich GmbH, Jülich, Germany.
    \\ $^{\ddagger}$ Computational Systems Biotechnology (AVT.CSB), RWTH Aachen University, Aachen, Germany.
}

\makeatletter
\def\blfootnote{\gdef\@thefnmark{}\@footnotetext}
\makeatother

\begin{document}

\maketitle

\begin{abstract}
Live-cell microscopy allows to go beyond measuring average features of cellular populations
to observe, quantify and explain biological heterogeneity.
Deep Learning-based instance segmentation and cell tracking form the gold standard analysis tools to
process the microscopy data collected, but tracking in particular suffers severely from low temporal resolution.
In this work, we show that approximating cell cycle time distributions in microbial colonies of \emph{C. glutamicum}
is possible without performing tracking, even at low temporal resolution.
To this end, we infer the parameters of a stochastic multi-stage birth process model
using the Bayesian Synthetic Likelihood method at varying temporal resolutions
by subsampling microscopy sequences, for which ground truth tracking is available.
Our results indicate, that the proposed approach yields high quality approximations
even at very low temporal resolution, where tracking fails to yield reasonable results.
\end{abstract}

\begin{keywords}
    Cell cycle times, stochastic simulation, live-cell microscopy, single-cell analysis, Bayesian inference
\end{keywords}

\blfootnote{\scriptsize Corresponding author: \href{mailto:k.noeh@fz-juelich.de}{\url{k.noeh@fz-juelich.de}}}

\setlength{\abovedisplayskip}{2pt}
\setlength{\belowdisplayskip}{2pt}

\section{Introduction}

Modern live-cell microscopy using microfluidic devices as lab-on-chip systems for massive parallel cultivation
of microbial colonies produces large amounts of image sequence at high spatio-temporal resolution.
This data allows to reveal biological heterogeneity at single-cell resolution within microbial populations \cite{grunberger_spatiotemporal_2015},
however efficient automated data analysis pipelines are crucial, as manual investigation quickly becomes infeasible.
The gold-standard analysis pipeline works by first detecting cell instances, 
nowadays using Convolutional Neural Networks (CNNs) \cite{maska_benchmark_2014,cutler_omnipose_2022}, 
which are then tracked over time \cite{ruzaeva_cell_2022,loffler_graph-based_2021,theorell_when_2019}. 
Cell tracking, which takes cell division into account, enables construction of so-called lineage trees \cite{theorell_when_2019}, 
which in turn enables the computation of cellular features like cell cycle times (CCTs)
for each individual cell and, thus, to compute distributions of such cellular features across populations.

Unfortunately, cell tracking is highly sensitive to 
temporal resolution of the data \cite{ruzaeva_cell_2022},
however temporal resolution is often traded off against throughput by increasing the number of parallel cultivations one
decides to observe.
Luckily, some important quantities of interest, like the average CCT, can be computed without tracking, 
though clearly at the cost of losing the single-cell characterization.
In the most straight-forward approach, an exponential curve is fitted against the number of detections over time,
however, this does not properly account for biological variability of CCTs within the population, as it only describes the mean behavior.
Further, the exponential curve is also known to be a good fit for the mean population behavior only in the 
limit of large populations.
Yet, microfluidic single-cell analysis considers just rather small populations, often descending from a single 
initial cell, often with less than 100 individuals.
At this scale, the stochasticity in cell division will govern the process.

Within this work, we show that using a stochastic multi-stage model with Erlang-distributed CCTs,
as proposed in \cite{kendall_role_1948} and \cite{yates_multi-stage_2017},
is able to approximate the distribution 
of CCTs within microbial populations without performing tracking. 
In particular, we demonstrate experimentally, that the proposed approach is more robust than tracking against 
decreases in temporal resolution.
For increased reliability of our analysis, 
we perform Bayesian inference using the Bayesian Synthetic Likelihood method \cite{price_bayesian_2018} for
quantifying parameter and predictive uncertainty.
We use a publicly available dataset \cite{seiffarth_obiwan-microbi_2022,seiffarth_data_2022},
which admits very high temporal resolution of \emph{C. glutamicum} cultivations under ideal growth conditions.
By subsampling the image sequences, we simulate lower temporal resolutions.
Our code is publicly available\footnote{\scriptsize\mbox{\url{https://jugit.fz-juelich.de/ias-8/multi-stage-growth}}}.

\section{Exponential Growth Model}
\label{sec:preliminaries}

Live-cell microscopy experiments investigating microbial growth yield temporally ordered image sequences
\( I = (I_1, \ldots, I_T ) \).
Using CNNs, each image gets processed and numbers of individual cell instances 
\( \obs = (Y_1, \ldots, Y_T )^\top \in \N^{T} \) 
for every frame of the sequence 
are identified.
The mean CCT $\cct$ is often estimated by performing exponential regression, 
i.e.~we choose 
\begin{align} 
    \label{eq:exp-model}
    \expreg_{\params}(t) = \n0 2^{\lambda t} 
    \numberthis
\end{align}
with $\lambda:=1/\cct$ and $\n0$ the initial population size. 
We then solve for \( \params := (\log_2 \n0, \lambda ) \in \N \times \R \) 
by minimizing the sum of squares residual
in the log-space,
which is solved by the ordinary least squares solution 
\begin{align} 
    \label{eq:ols}
    \params^* := (\design^\top\!\design)^{-1} \design^\top\!\tilde{\obs}, 
    \numberthis
\end{align}
where \( \design := ((1, 1), \ldots, (1, T))^\top \in \R^{T \times 2} \) is the design matrix and  
\( \tilde{\obs} := (\log_2 Y_1, \ldots, \log_2 Y_T)^\top \) are the observed population sizes at time $t$ in log-space.
Note the choice of base 2 in \Cref{eq:exp-model},
which accounts for the doubling of cells through cell division.

\subsection{Memoryless birth process}

The doubling induced by cell division can also be represented by considering a simple stochastic 
birth model, where each cell divides after some time $\tc$, the CCT, into two daughters.
We denote the current population as $\Nexp(t) \in \N$.
The CCT $\tc$ any individual in $\Nexp(t)$
takes until division is assumed to be exponentially distributed at rate \( \lambda = 1 / \cct \).
Lending notation from chemical reactions we denote this process as
\begin{align}
    X \overset{\lambda}{\to} 2 X,
\end{align}
for any individual \( X \) from the current population $\Nexp(t)$.
It was shown \cite{feller_grundlagen_1939}, that the expected population size \( \E[\Nexp(t)] \) 
is exactly the exponential curve from \Cref{eq:exp-model}, i.e.~
\begin{align}
    \E[\Nexp(t)] = \n0 2^{ \lambda t } = g_{\params}(t).
\end{align}
Therefore, the exponential curve only describes the mean population growth.
Investigating the exponential distribution that we assumed on the CCT $\tc$,
we observe that it is not capable to capture empirically observed CCT distributions correctly
(c.f.~\Cref{fig:simemp}, where $k=1$) as in \cite{kendall_role_1948,yates_multi-stage_2017}.

\section{Multi-Stage Growth Model}
\label{sec:multi-stage}

A more appropriate model for cell division was proposed in~\cite{kendall_role_1948},
assuming that a cell cycle consists of $k \in \N$ stages, through which the cell has to live before it can
subdivide.
By assuming that the time before transitioning to the next stage is exponentially distributed at rate 
$\lambda k$, the sum of $k$ transitions takes on average $\cct = 1/\lambda$ units of time \cite{yates_multi-stage_2017}.
Similar to \Cref{eq:exp-model}, this process resembles a chain of $k$ reactions
\begin{align}
    \label{eq:multi-stage}
    X_1 \overset{\lambda k}{\to} 
    X_2 \overset{\lambda k}{\to}
    \ldots \overset{\lambda k}{\to} 
    X_k \overset{\lambda k}{\to} 
    2 X_1,
\end{align}
where $X_j$ is an individual from the population in stage $j \in \{1, \ldots, k\}$.
In distinction to the memoryless process $\Nexp(t)$, 
we refer to this new multi-stage process as 
\begin{align}
    \Nmulti(t) = \sum_{j=1}^k \Nmulti_j(t),
\end{align}
being the sum of current individuals $\Nmulti_j(t)$ over all $k$ stages
and which is parametrized by $\mparams = (\n0, \lambda, k) \in \N \times \R \times \N$.

\begin{figure}
    \centering
    {\includegraphics[width=.9\columnwidth]{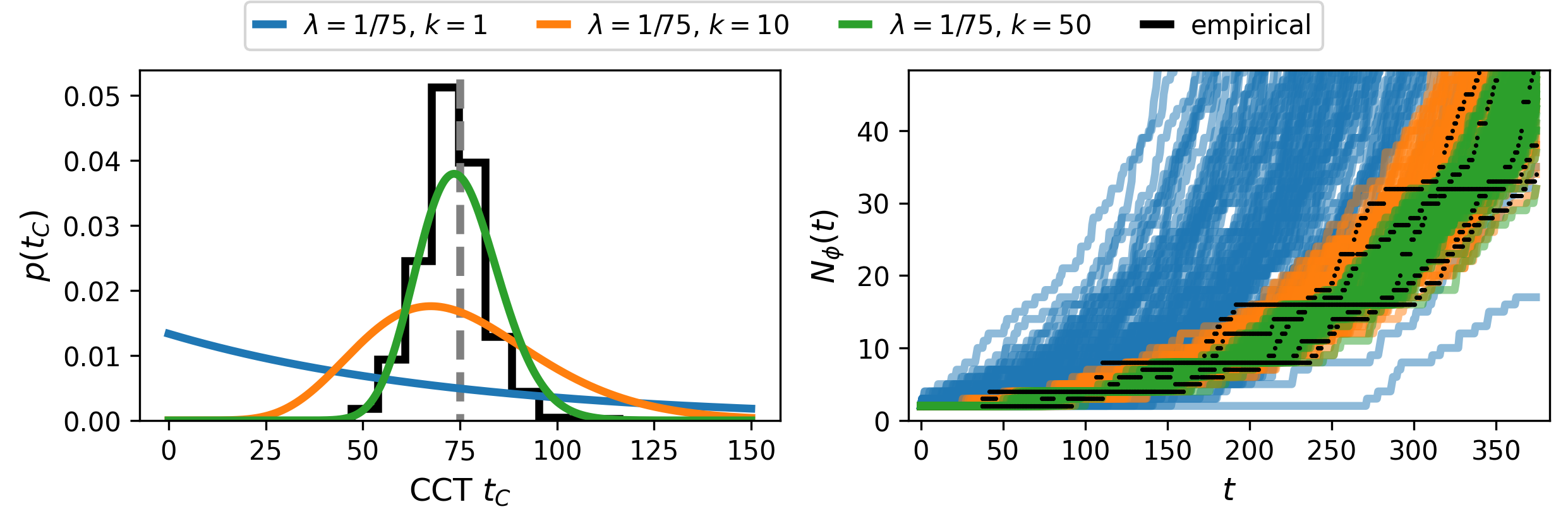}}
    \vspace{-1.\baselineskip}
    \caption{\small Comparison between empirical distribution of CCTs obtained from \cite{seiffarth_data_2022}
        and those obtained from the memoryless and multi-stage birth process models on the left.
        Forward simulations of the respective stochastic processes on the right, 
        with black dots indicating number of cells per frame for all sequences.
    }
    \label{fig:simemp}
\end{figure}

For this model, the CCT $\tc$ of the multi-stage process $\Nmulti(t)$ is Erlang-distributed, 
i.e.~$\tc \sim \mathrm{Erlang}(\alpha, \beta)$, 
with shape parameter $\alpha=k$ and rate parameter $\beta = \lambda k$~\cite{yates_multi-stage_2017}.
By considering the mean \( \alpha / \beta = 1 / \lambda \)  and variance \( \alpha / \beta^2 = 1 / (\lambda^2 k) \) of the 
Erlang distribution, we observe the influence of $k$ in controlling the variance of the CCT distribution.
For $k=1$, this model recovers the exponential model from \Cref{eq:exp-model}, while in the limit of 
$k \to \infty$, the CCT distribution approaches a Dirac delta at $\cct$,
meaning that all cells divide after exactly $\cct$ units of time.
We illustrate different densities of CCT distributions and corresponding simulation results of 
$\Nmulti(t)$ for varying numbers of cell stages $k$ in 
\Cref{fig:simemp}.

\vspace{-5mm}

\subsection{Statistical inference for multi-stage growth}
\label{sec:inference}

To infer CCTs from observed cell counts $\obs$,
the task is to estimate the parameters \( \mparams = (\n0, \lambda, k) \).
An intuitive approach is maximizing the likelihood $\P(\Nmulti(t) = Y_t|\mparams)$ of observing the data,
i.e.~estimate $\mparams$ as
\begin{align}
    \label{eq:ex-likeli}
    \mparams^*  &= \argmax_{\mparams} \prod_{t=1}^T \P(\Nmulti(t) = Y_t|\mparams),
    \numberthis
\end{align}
given observed cell counts $\obs$.
However, the likelihood $\P(\Nmulti(t)=Y_t|\mparams)$ cannot be computed analytically and estimating it using Monte-Carlo integration
suffers severely from truncation errors.
Therefore, we approximate it by moment matching a normal distribution 
$\normal( \E[\Nmulti(t)], \Var[\Nmulti(t)] )$,
in which case we obtain an approximate likelihood
\begin{align}
    \label{eq:approx-likeli}
    \likeli_{\mparams}(\obs | \mparams) = \prod_{t=1}^T \normal\!\left( Y_t \, | \, \E[\Nmulti(t)],
        \Var[\Nmulti(t)] \right).
    \numberthis
\end{align}
Unfortunately, the multi-stage model $\Nmulti(t)$ admits analytic solutions for the mean $\E[\Nmulti(t)]$ and 
variance $\Var[\Nmulti(t)]$ of the population size only for specific numbers of stages 
$k$~\cite{kendall_role_1948, yates_multi-stage_2017}.
Thus, by simulating $M$ trajectories $\tNmulti{(i)}(t)$ from the process $\Nmulti(t)$ 
we resort to Monte-Carlo estimates for mean
\begin{align}
    \mNmulti(t) := \frac{1}{M} \sum_{i=1}^M \tNmulti{(i)}(t) 
\end{align}
and variance
\begin{align}
    \vNmulti(t) := \frac{1}{M-1} \sum_{i=1}^M (\tNmulti{(i)}(t) - \mNmulti(t))^2 .
\end{align}
Given these estimates, we approximate the likelihood from \Cref{eq:approx-likeli} as
\begin{align}
    \hat{\likeli}_{\mparams}(\obs | \mparams) = \prod_{t=1}^T \normal\!\left( Y_t \, | \, \mNmulti(t), \vNmulti(t) \right),
\end{align}
which is computationally tractable, 
since simulation of $\Nmulti(t)$ can be performed using e.g.~the $\tau$-leaping algorithm \cite{gillespie_approximate_2001}.

Apart from the likelihood function, which relates data and model parameters, 
some prior knowledge is available for determining the model parameters.
Excluding the possibility of cell death, we restrict $\lambda \in \R^+$,
in which case the multi-stage process $\Nmulti(t)$ is monotonically increasing,
motivating the restriction $\n0 \sim p(\n0) = \U_{[1, Y_0]}$.
In practice, exponential regression, as in \Cref{eq:ols}, gives reliable estimates for the mean CCT $\cct$.
We incorporate this faith by restricting the prior on $1/\lambda = \cct \sim \U_{[0.5\cdot\cct, 2\cdot\cct]}$,
resulting in a reciprocal distribution as prior $p(\lambda)$.
For $k$, we choose a log-uniform prior $p(k)$, i.e.~$\log k \sim \U_{[0, 5]}$.
Moreover, we consider a further parameter $t_0$ which serves as an offset between simulation beginning and
first observation $Y_0$, i.e.~we shift all $Y_t$ to $Y_{t_0 + t}$. 
As the internal states of the initial cells $n_0$ within the cascade of stages from \Cref{eq:multi-stage} 
is unknown, adding this offset allows for some flexiblity to account for this lack of knowledge.
In order to avoid non-identifiabilities from strong correlations between $\n0$ and $t_0$, we further
impose the prior $t_0 \sim p(t_0) = \U_{[0, 2/\lambda]}$, limiting the offset to at most twice the CCT.

This prior knowledge is implemented into our parameter estimation by taking a Bayesian perspective
on the problem, i.e.~we consider the approximate posterior distribution 
\begin{align}
    \label{eq:posterior}
    \hat{\pi}( \mparams | \obs ) \propto \hat{\likeli}_{\mparams}( \obs | \mparams ) \cdot p( \mparams ),
\end{align}
where $p( \mparams ) = p( \lambda ) \cdot p( k ) \cdot p( \n0 ) \cdot p( t_0 )$.
Performing Bayesian inference on this approximate posterior \( \hat{\pi}( \mparams | \obs ) \) using e.g.~
Markov chain Monte Carlo (MCMC) sampling is also known as Bayesian Synthetic Likelihood method \cite{price_bayesian_2018}.

\vspace{-.5mm}

\section{Experiments}
\label{sec:experiments}

We perform Bayesian inference using MCMC sampling on the approximate posterior distribution from \Cref{eq:posterior}
to estimate multi-stage model parameters
for each of the five image sequences from \cite{seiffarth_data_2022} for 18 subsampling factors
between 1 and 40, meaning we omit every $n$th image, which simulates low temporal resolution during image acquisition.
We call this experiment \multistage.
As a result, we obtain parameter estimates, which can be used to compute an Erlang distribution
as an approximation to the CCT distribution as in \Cref{sec:multi-stage}.

In order to compare the CCT distributions obtained from \multistage{} against those computed from tracking, 
we also perform tracking on the same subsamples of the five image sequences as in \multistage.
We choose two algorithms, \KITGE{} \cite{loffler_graph-based_2021} and \MUCZ{}
from the Cell Tracking Challenge \cite{maska_benchmark_2014}, and a third one, \AT{} \cite{ruzaeva_cell_2022}, 
which shows improved performance over \KITGE{} in the original publication.

\begin{figure}
    \centering
    \includegraphics[width=.9\columnwidth]{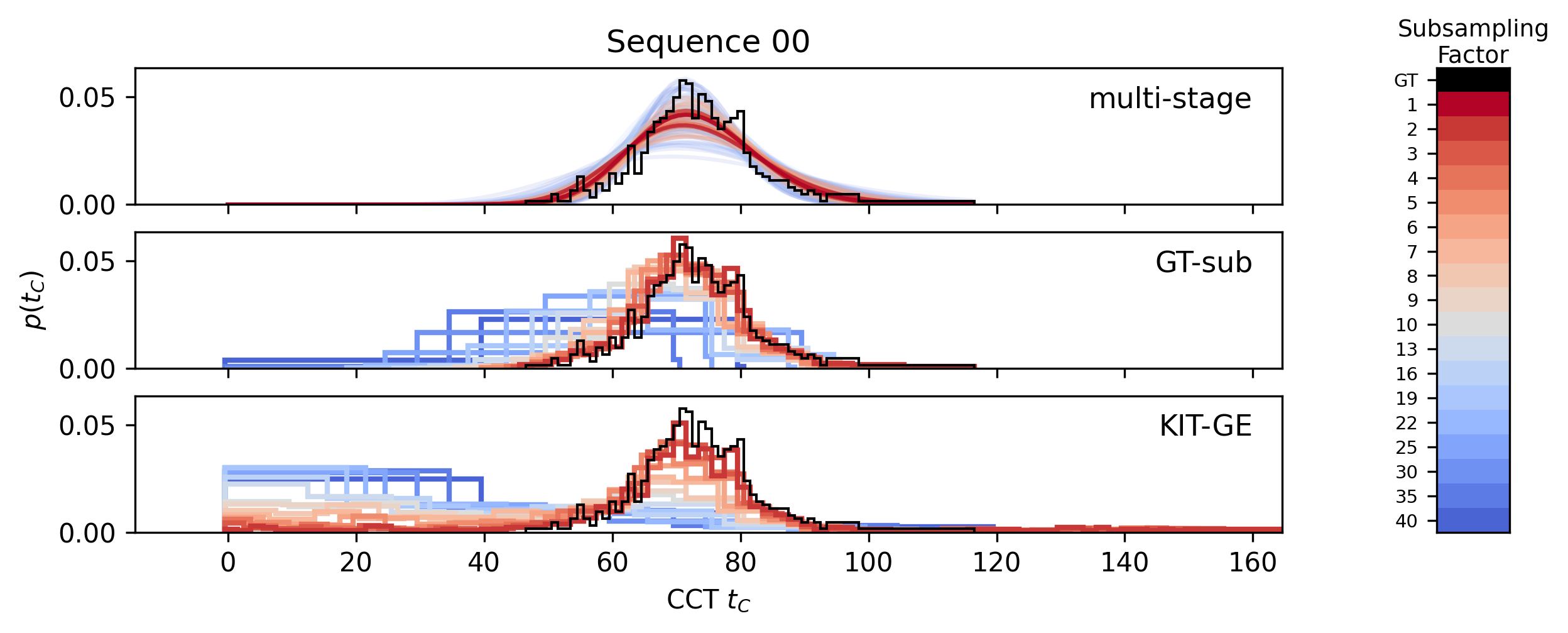}
    \vspace{-1.\baselineskip}
    \caption{\small Exemplary visualization of CCT distributions for the first sequence from \cite{seiffarth_data_2022} computed with 
        \multistage, \KITGE{} and \GTsub{} (c.f.~\Cref{sec:experiments}) at various subsampling factors.}
    \label{fig:results-seq}
\end{figure}

Further, we also consider the distributions of CCTs obtained from subsampling the GT tracking,
i.e.~the CCT distributions computed, if we could solve the tracking exactly under subsampling.
This experiment we call \GTsub, which should give an upper bound on the quality of estimation of 
CCT distributions from \tracking.
The different distributions obtained
using the different approaches mentioned are depicted exemplary for one sequence in \Cref{fig:results-seq}.

\subsection{Technical details}
\label{sec:tech-details}

Multi-stage experiments were run on six machines, each equipped with two AMD EPYC 7742, 2.25 GHz processors,
running Rocky Linux 8.
For each sequence and subsampling combination, 4 Metropolis chains were simulated using the sampling software
\texttt{hopsy} \cite{jadebeck_hops_2021, paul_hopsy_2023} and a custom-implemented \texttt{C++} simulator for simulating the
stochastic multi-stage birth process using a $\tau$-leaping-like algorithm \cite{gillespie_approximate_2001}.
The latter simulations are parallelized using \texttt{OpenMP} and were run on 20 threads each.
For every evaluation of \Cref{eq:approx-likeli}, we drew $M=80$ samples.
Chains were run either for up to 50~000 iterations with a thinning factor of 20 or until convergence 
was diagnosted using the $\hat{R}$-diagnostics
\cite{gelman_inference_1992}, where we set the convergence threshold to 1.1.
As a proposal algorithm, we used a Gaussian distribution with standard deviation 0.5.
The resulting Metropolis chain moves in continuous space, 
but since some of our parameters in \Cref{sec:multi-stage} are discrete,
we round continuous values to the closest integer.

Tracking was computed on a single machine with two AMD EPYC 7282, 2.8 GHz processors, running Ubuntu 22.04.
Implementations for \KITGE{} and \MUCZ{} were taken from the Cell Tracking Challenge 
webpage\footnote{\scriptsize{\url{http://celltrackingchallenge.net/participants/}}},
and from the original implementation\footnote{\scriptsize{\url{https://github.com/kruzaeva/activity-cell-tracking}}} 
for \AT.

\section{Results \& Discussion}
\label{sec:results}

\begin{figure}
    \centering
    \includegraphics[width=.9\columnwidth]{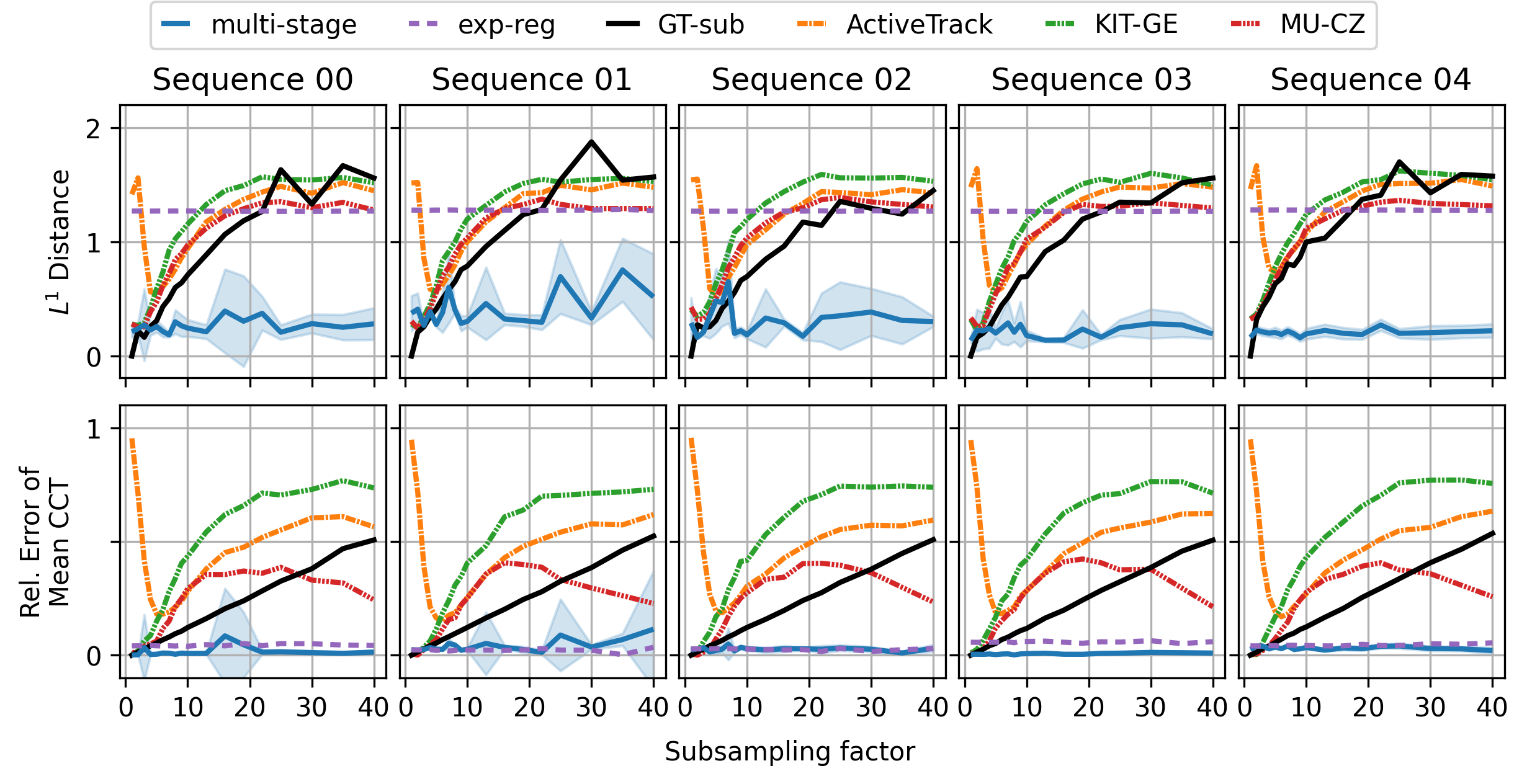}
    \vspace{-1.\baselineskip}
    \caption{\small Comparison of approximation quality in terms of \Lp distance (top) and relative error of the mean (bottom)
        between GT CCT distribution and CCT distributions computed using exponential regression (\texttt{exp-reg}),
        the multi-stage model approach \multistage, tracking on subsampled data (\tracking) and subsampling of the
        GT tracking, \GTsub. 
        Note that for exponential regression, we assume CCTs to follow an exponential distribution.
        Blue shaded area indicates standard deviation from the mean in the Monte-Carlo estimates obtained from
        posterior parameter samples.
        In both cases, lower is better.
    }
    \label{fig:results-kl}
\end{figure}

In \Cref{fig:results-kl}, we report the \Lp distance and absolute errors of the mean between
the GT CCT distribution and those obtained from the five approaches \multistage, \tracking, as well as \GTsub{}
in order to quantify approximation quality.
For all five sequences, we observe that \multistage{} is capable to 
approximate empirical cell cycle time distribution very well, even for high subsampling factors,
yielding almost constant \Lp distance and low relative error of the mean CCT.
\tracking{} as well as \GTsub{} show 
increasing \Lp distance and absolute error in the estimated mean CCT as subsampling increases.
In fact, already for a subsampling factor of 10, which corresponds to a temporal resolution of 1 frame every 10 minutes or approximately
1/7 of the mean GT CCT, \multistage{} is on par with \GTsub{} in terms of \Lp distance.
Against \tracking, \multistage{} yields better results already
at higher temporal resolutions of approximately 5 frames.
However, the mean CCTs obtained from \GTsub{} as well as those from \tracking{} 
quickly diverge from the true mean, as can be seen from the relative error of mean CCT in \Cref{fig:results-kl}.
On the other hand, we observe a slight bias of the mean CCT estimates from \texttt{exp-reg} and \multistage, 
which is even present at highest temporal resolution.

Interestingly, we observe only slight to no increments in the predictive uncertainty of the \Lp distance
and relative error computed for \multistage, which contradicts the expectation that less data should
lead to higher parameter uncertainty.
Investigation of the parameter posterior shows a highly multi-modal density with large equi-probable
regions stemming from the discrete parameter space.
We assume the usage of a continuous sampling algorithm (c.f. \Cref{sec:tech-details}) on such a complex posterior 
distribution to be a cause for impaired and possibly miss-diagnosed convergence,
which may lead to underestimation of predictive uncertainty.
We hypothesize that using a custom sampling algorithm combined with parallel tempering
techniques \cite{swendsen_replica_1986} may greatly improve the speed and convergence of MCMC sampling.

Overall, our results suggest, that a multi-stage birth process model is capable to approximate
the CCT distribution much more robustly than tracking algorithms, in particular when temporal resolution is low.
We hypothesize that for the image data at hand, which exhibits storybook exponential microbial growth,
the multi-stage process is a very appropriate model, similar to exponential regression.
However, unlike exponential regression, the multi-stage model is capable to
capture intra-population variance in CCTs.
We emphasize this observation by reporting \Lp distance and relative error of the mean for
the CCT distribution obtained using exponential regression (\Cref{eq:ols}) in \Cref{fig:results-kl} denoted as \texttt{exp-reg}.
Here, we assume the exponential regression to be the exact solution of the multi-stage model
for fixed $k=1$, yielding an exponentially distributed CCT.

Nevertheless, our approach comes at the cost of only obtaining an approximate result
for population-level CCT distributions,
whereas tracking yields much more fine-grained information.
Further, similar to exponential regression, our approach assumes
exponential population growth and time-homogeneous CCT distributions,
both which may not always be the case when working with real-world data.
Yet, we show that obtaining such approximations at high quality is possible, even
when tracking is not applicable anymore.

\section{Conclusion}
\label{sec:conclusion}

Within this work, we show that the high correspondance between a multi-stage model and microbial growth
can be leveraged to approximate CCT distributions from live-cell microscopy image sequences even at low temporal
resolutions, where cell tracking, the gold standard analysis tool for single-cell image analysis, fails
to yield reasonable results.
Although the assumption on ideal growth conditions might be rather restrictive,
we envision our approach to still be helpful especially in screening studies, where
throughput is preferred over high temporal resolution.
As such, this work bridges the gap between coarse approximations of CCTs computed with exponential regression
and single-cell resolved CCT distributions obtained from tracking.

\section{Compliance with Ethical Standards}

This research study was conducted retrospectively
using microbial data made available in open
access by \cite{seiffarth_data_2022}. Ethical approval was
not required.

\section{Acknowledgements}

This work was performed as part of the Helmholtz School for Data
Science in Life, Earth and Energy (HDS-LEE) and received funding from
the Helmholtz Association.
This work was supported by the President's Initiative and Networking Funds of the Helmholtz Association of German Research Centres [EMSIG ZT-I-PF-04-044].
The authors gratefully acknowledge computing time on the supercomputer JUSUF \cite{vieth_jusuf_2021} at 
Forschungszentrum Jülich under grant no. \texttt{paj2312}.

\bibliographystyle{IEEEbib}
\bibliography{literature}

\begin{thebibliography}{10}

\bibitem{grunberger_spatiotemporal_2015}
Alexander Grünberger, Christopher Probst, Stefan Helfrich, Arun Nanda, Birgit
  Stute, Wolfgang Wiechert, Eric Von~Lieres, Katharina Nöh, Julia Frunzke, and
  Dietrich Kohlheyer,
\newblock ``Spatiotemporal microbial single‐cell analysis using a
  high‐throughput microfluidics cultivation platform,''
\newblock {\em Cytometry Part A}, vol. 87, no. 12, pp. 1101--1115, Dec. 2015.

\bibitem{maska_benchmark_2014}
Martin Maška, Vladimír Ulman, David Svoboda, Pavel Matula, Petr Matula,
  Cristina Ederra, Ainhoa Urbiola, Tomás España, Subramanian Venkatesan,
  Deepak~M.W. Balak, Pavel Karas, Tereza Bolcková, Markéta Štreitová, Craig
  Carthel, Stefano Coraluppi, Nathalie Harder, Karl Rohr, Klas E.~G. Magnusson,
  Joakim Jaldén, Helen~M. Blau, Oleh Dzyubachyk, Pavel Křížek, Guy~M.
  Hagen, David Pastor-Escuredo, Daniel Jimenez-Carretero, Maria~J.
  Ledesma-Carbayo, Arrate Muñoz-Barrutia, Erik Meijering, Michal Kozubek, and
  Carlos Ortiz-de Solorzano,
\newblock ``A benchmark for comparison of cell tracking algorithms,''
\newblock {\em Bioinformatics}, vol. 30, no. 11, pp. 1609--1617, June 2014.

\bibitem{cutler_omnipose_2022}
Kevin~J. Cutler, Carsen Stringer, Teresa~W. Lo, Luca Rappez, Nicholas
  Stroustrup, S.~Brook~Peterson, Paul~A. Wiggins, and Joseph~D. Mougous,
\newblock ``Omnipose: a high-precision morphology-independent solution for
  bacterial cell segmentation,''
\newblock {\em Nature Methods}, vol. 19, no. 11, pp. 1438--1448, Nov 2022.

\bibitem{ruzaeva_cell_2022}
Karina Ruzaeva, Jan-Christopher Cohrs, Keitaro Kasahara, Dietrich Kohlheyer,
  Katharina Nöh, and Benjamin Berkels,
\newblock ``Cell tracking for live-cell microscopy using an
  activity-prioritized assignment strategy,''
\newblock in {\em 2022 IEEE 5th International Conference on Image Processing
  Applications and Systems (IPAS)}, 2022, vol. Five, pp. 1--7.

\bibitem{loffler_graph-based_2021}
Katharina Löffler, Tim Scherr, and Ralf Mikut,
\newblock ``A graph-based cell tracking algorithm with few manually tunable
  parameters and automated segmentation error correction,''
\newblock {\em PLOS ONE}, vol. 16, no. 9, pp. 0249257, 2021.

\bibitem{theorell_when_2019}
Axel Theorell, Johannes Seiffarth, Alexander Grünberger, and Katharina Nöh,
\newblock ``When a single lineage is not enough: {Uncertainty}-{Aware}
  {Tracking} for spatio-temporal live-cell image analysis,''
\newblock {\em Bioinformatics}, vol. 35, no. 7, pp. 1221--1228, Apr. 2019.

\bibitem{kendall_role_1948}
David~G. Kendall,
\newblock ``On the {Role} of {Variable} {Generation} {Time} in the
  {Development} of a {Stochastic} {Birth} {Process},''
\newblock {\em Biometrika}, vol. 35, no. 3/4, pp. 316, Dec. 1948.

\bibitem{yates_multi-stage_2017}
Christian~A. Yates, Matthew~J. Ford, and Richard~L. Mort,
\newblock ``A {Multi}-stage {Representation} of {Cell} {Proliferation} as a
  {Markov} {Process},''
\newblock {\em Bulletin of Mathematical Biology}, vol. 79, no. 12, pp.
  2905--2928, Dec. 2017.

\bibitem{price_bayesian_2018}
L.~F. Price, C.~C. Drovandi, A.~Lee, and D.~J. Nott,
\newblock ``Bayesian {Synthetic} {Likelihood},''
\newblock {\em Journal of Computational and Graphical Statistics}, vol. 27, no.
  1, pp. 1--11, Jan. 2018.

\bibitem{seiffarth_obiwan-microbi_2022}
Johannes Seiffarth, Tim Scherr, Bastian Wollenhaupt, Oliver Neumann, Hanno
  Scharr, Dietrich Kohlheyer, Ralf Mikut, and Katharina N\"{o}h,
\newblock ``Obiwan-microbi: Omero-based integrated workflow for annotating
  microbes in the cloud,''
\newblock {\em SoftwareX}, vol. 26, pp. 101638, May 2024.

\bibitem{seiffarth_data_2022}
Johannes Seiffarth, Luisa Blöbaum, Katharina Löffler, Tim Scherr, Alexander
  Grünberger, Hanno Scharr, Ralf Mikut, and Katharina Nöh,
\newblock ``{Data for - Tracking one in a million: Performance of automated
  tracking on a large-scale microbial data set},''
\newblock {\em Zenodo}, Oct. 2022.

\bibitem{feller_grundlagen_1939}
Willy Feller,
\newblock ``Die {Grundlagen} der {Volterraschen} {Theorie} des {Kampfes} ums
  {Dasein} in wahrscheinlichkeitstheoretischer {Behandlung},''
\newblock {\em Acta Biotheoretica}, vol. 5, no. 1, pp. 11--40, May 1939.

\bibitem{gillespie_approximate_2001}
Daniel~T. Gillespie,
\newblock ``Approximate accelerated stochastic simulation of chemically
  reacting systems,''
\newblock {\em The Journal of Chemical Physics}, vol. 115, no. 4, pp.
  1716--1733, July 2001.

\bibitem{jadebeck_hops_2021}
Johann~F Jadebeck, Axel Theorell, Samuel Leweke, and Katharina Nöh,
\newblock ``{HOPS}: high-performance library for (non-)uniform sampling of
  convex-constrained models,''
\newblock {\em Bioinformatics}, vol. 37, no. 12, pp. 1776--1777, July 2021.

\bibitem{paul_hopsy_2023}
Richard~D. Paul, Johann~F. Jadebeck, Anton Stratmann, Wolfgang Wiechert, and
  Katharina N{\"o}h,
\newblock ``hopsy - a methods marketplace for convex polytope sampling in
  python,''
\newblock {\em bioRxiv}, 2023.

\bibitem{gelman_inference_1992}
Andrew Gelman and Donald~B. Rubin,
\newblock ``Inference from {Iterative} {Simulation} {Using} {Multiple}
  {Sequences},''
\newblock {\em Statistical Science}, vol. 7, no. 4, pp. 457--472, Nov. 1992,
\newblock Publisher: Institute of Mathematical Statistics.

\bibitem{swendsen_replica_1986}
Robert~H. Swendsen and Jian-Sheng Wang,
\newblock ``Replica {Monte} {Carlo} {Simulation} of {Spin}-{Glasses},''
\newblock {\em Physical Review Letters}, vol. 57, no. 21, pp. 2607--2609, Nov.
  1986.

\bibitem{vieth_jusuf_2021}
Benedikt von~St Vieth,
\newblock ``{JUSUF}: {Modular} {Tier}-2 {Supercomputing} and {Cloud}
  {Infrastructure} at {Jülich} {Supercomputing} {Centre},''
\newblock {\em Journal of large-scale research facilities JLSRF}, vol. 7, pp.
  A179--A179, Oct. 2021.

\end{thebibliography}

\typeout{get arXiv to do 4 passes: Label(s) may have changed. Rerun}
\end{document}